\documentclass[prd,showpacs,showkeys,preprint]{revtex4}

\usepackage{amsmath}
\usepackage{amssymb}
\usepackage{yfonts}[1988/10/03]
\usepackage{graphicx}
\newcommand {\beq}{\begin{equation}}
\newcommand {\eeq}{\end{equation}}
\newcommand {\beqa}{\begin{eqnarray}}
\newcommand {\eeqa}{\end{eqnarray}}

\begin{document}
\title{Detection  relic gravitational waves in thermal case by using Adv.LIGO data of GW150914 }
\author{ Basem Ghayour$^{1}$\footnote{ba.ghayour@gmail.com},  Jafar Khodagholizadeh $^{2}$\footnote{gholizadeh@ipm.ir}}
\affiliation{$^{1}$ School of Physics, University of Hyderabad,
Hyderabad-500 046. India.\\
$^{2}$ Farhangian University, P. O. Box 11876-13311,Tehran, Iran.}

\date{\today} 

\begin{abstract}
The thermal spectrum of relic gravitational waves causes the  new amplitude that called `modified amplitude'. Our analysis shows that, there exist   some chances for  detection of the thermal spectrum  in addition to the usual spectrum by  using Adv.LIGO data of GW150914 and detector based on the maser light(Dml).  The behaviour of the inflation and reheating stages are often known as power law expansion like $S(\eta)\propto \eta^{1+\beta}$, $S(\eta)\propto \eta^{1+\beta_s}$ respectively. The $\beta$ and $\beta_s$ have an unique effect on the shape of the spectrum. We find some upper bounds on the $\beta$ and $\beta_s$ by comparison the  usual and thermal spectrum with  the Adv.LIGO and  Dml. As this result modifies our information about the nature of the evolution of inflation and reheating stages.
\end{abstract}
\maketitle
\section{\label{sec:level1}Introduction }
The relic gravitational waves (RGWs) have a  wide range  of frequency $\sim($10$^{-19}$Hz -10$^{11}$Hz). They are generated before and during inflation stage. They did not have interaction with  other matters during their travel from the early universe until now. Therefore they contain valuable information about the early universe. Thus we can obtain  the information by detecting RGWs on the through  range  of the frequency. Nowadays the people are trying to detect the waves at different frequency ranges like: $\sim($10$^{-19}$Hz -10$^{-16}$Hz) by Planck \cite{1}, $\sim($10$^{-7}$Hz -10$^{0}$Hz) by eLISA \cite{2}, $\sim($10$^{-1}$Hz -10$^{4}$Hz) by Advanced.LIGO (Adv.LIGO) \cite{3}, $\sim($10$^{0}$Hz -10$^{4}$Hz) by Einstein telescope (ET) \cite{4}, GHz band by detector based on the maser light (Dml)\cite{5} and etc.  It is believed that thermal spectrum of RGWs exists from a pre-inflationary stage   and it may affect on the CMB temperature and the polarization anisotropies   in the  low frequency range  ($\sim$10$^{-18}$Hz-10$^{-16}$Hz) \cite{6}. Then thermal spectrum of RGWs extend this investigation to the general pre-inflationary scenario by assuming the effective equation of state, $ \omega $ being a free parameter \cite{61}. Also in high  frequency range  $\sim($10$^{8}$Hz -10$^{11}$Hz), extra dimensions cause  thermal gravitational waves (or, equivalently, a primordial background of gravitons) \cite{7}. For more details about the extra dimension see Ref.\cite{23}.  For the gravity-wave background origin, any fit of the CMB anisotropy in term of gravity background should include a thermal dependence in the spectrum \cite{71}. Thus in the middle range $\sim($10$^{-16}$Hz -10$^{8}$Hz), this thermal spectrum causes the  new amplitude that called `modified amplitude'    \cite{8}.   We have analysed the results of  modified amplitude by comparison it with the sensitivity of the Adv.LIGO, ET and LISA in  \cite{8}. Recently Adv.LIGO has detected the effect of waves  of
a pair of black holes called GW150914 \cite{9} with a peak gravitational-wave strain of $1.0\times10^{-21}$ in the frequency range 35 to 250 Hz. There is an average measured
sensitivity  in the range $\sim($10$^{-1}$Hz -10$^{4}$Hz) of the Adv.LIGO detectors (Hanford and Livingston) during the
time analysed to determine the significance of GW150914 (Sept 12 - Oct 20, 2015) \cite{3}. Therefore in this work, we upgrade our previous work \cite{8} by comparison the thermal spectrum with average measured
sensitivity of Adv.LIGO and Dml. We show that there are some chances for detection the spectrum of  RGWs in usual and thermal case.

 On the other hand the different stages of the evolution of the universe (inflation, reheating, radiation, matter and acceleration) cause some variation in the shape of the spectrum of the RGWs.
 The behaviour of the inflation and reheating stages are often known as power law expansion like $S(\eta)\propto \eta^{1+\beta}$, $S(\eta)\propto \eta^{1+\beta_s}$ respectively. The $S$ and $\eta$ are scale factor and conformal time respectively and $\beta, \beta_s$ constrained on the $1+\beta<0$ and  $1+\beta_s>0$ \cite{11,12}. The $\beta$ and $\beta_s$ have an unique effect on the shape of the spectrum in the full range $\sim($10$^{-19}$Hz -10$^{11}$Hz) and high frequency range $\sim($10$^{8}$Hz -10$^{11}$Hz) respectively. Therefore  these two parameters play main role in the spectrum of the RGWs. Thus, we are interested to obtain some upper bounds on the $\beta$ and $\beta_s$ by comparison the  usual and thermal spectrum with the average measured
sensitivity of the Adv.LIGO and  Dml. As obtained result of the upper bounds can modify our information about the nature of the evolution of inflation and reheating stages. In the present work, we use the unit $c=\hbar = k_{B} =1$.

 \section{The spectrum of gravitational waves}
The perturbed metric for a homogeneous  isotropic  flat Friedmann-Robertson-Walker universe  can be written   as
\begin{equation}
d s^{2}= S^{2}(\eta)(d\eta^{2}-(\delta_{ij}+h_{ij})dx^{i}dx^{j}),
\end{equation}
where $\delta_{ij}$ is the Kronecker delta  symbol.  The $h_{ij} $ are  metric perturbations  with  the transverse-traceless properties i.e; $\nabla_i h^{ij} =0, \delta^{ij} h_{ij}=0$.
 The
 gravitational waves are described with  the
 linearized field equation given by
 \begin{equation}\label{weq}
 \nabla_{\mu} \left( \sqrt{-g} \, \nabla^{\mu} h_{ij}(\bf{x}, \eta)\right)=0.
 \end{equation}
The tensor perturbations have two independent physical degrees of freedom  and are denote as $h^{+}$ and $h^{\times}$ that called polarization modes. We express $h^{+}$ and $h^{\times}$ in terms
of the creation ($a^{\dagger}$) and annihilation ($a$) operators,
\begin{eqnarray}\label{1}
\nonumber  h_{ij}({\bf x},\eta)=\frac{\sqrt{16\pi} l_{pl}}{S(\eta)} \sum_{\bf{p}} \int\frac{d^{3}k}{(2\pi)^{3/2}} {\epsilon}_{ij} ^{\bf {p}}(\bf {k}) \\
 \times  \frac{1}{\sqrt{2 k}} \Big[a_{\bf{k}}^{\bf {p}}h_{\bf {k}}^{\bf {p}}(\eta) e^{i \bf {k}.\bf {x}} +a^{\dagger}_{\bf {k}} {^{\bf {p}}} h^{*}_{\bf {k}}{^{\bf {p}}} (\eta)e^{-i\bf{k}.\bf{x}}\Big],
\end{eqnarray}
where  $\bf{k}$ is the comoving wave
number, $k=|\bf {k}|$, $l_{pl}= \sqrt{G}$ is the
Planck's length and $\bf{ p}= +, \times$ are polarization modes. The polarization tensors
$\epsilon_{ij} ^{{\bf p}}({\bf k})$ are symmetric and transverse-traceless  $ k^{i} \epsilon_{ij} ^{{\bf p}}({\bf k})=0, \delta^{ij} \epsilon_{ij} ^{{\bf p}}({\bf k})=0$ and
satisfy  the conditions $\epsilon^{ij {\bf p}}({\bf k})   \epsilon_{ij}^{{\bf p}^{\prime}}({\bf k})= 2  \delta_{ {\bf p}{{\bf p}}^{\prime}} $ and $ \epsilon^{{\bf p}}_{ij} ({\bf -k}) = \epsilon^{{\bf p}}_{ij} ({\bf k}) $. Also the creation and annihilation operators  satisfy
$[a_{{\bf k}}^{{\bf p}},a^{\dagger}_{{\bf k} ^{\prime}} {{^{{\bf p}}}^{\prime}}]= \delta_{{{\bf p}} {\bf {p}}^{\prime} }\delta^{3}({\bf k}-{{\bf k}}^{\prime})$ and the initial vacuum state is defined  as
\begin{equation}
a_{\bf{k}}^{\bf{p}}|0\rangle = 0,
\end{equation}
for each $\bf {k}$ and $\bf {p}$.
 For a fixed  wave number $\bf{k} $ and a fixed polarization state $\bf{p}$ the eq.(\ref{weq}) gives coupled Klein-Gordon equation as follows:
 \begin{equation}\label{zz}
f^{\prime \prime}_{k}+\Big(k^{2}-\frac{S^{\prime \prime}}{S} \Big)f_{k}=0.
\end{equation}
 where $h_{k}(\eta)=f_{k}(\eta)/S(\eta)$ \cite{11,12} and prime means derivative with respect to the conformal time. Since the  polarization states are  same, we consider  $f_{k}(\eta)$ without the polarization  index.
The   solution of the above equation for the different stages of the universe  are given in appendix.[A]. There is another state that called  `thermal vacuum state' see appendix.[B] for more details. The amplitude of the RGWs in  thermal vacuum state are as follows \cite{8}:

\begin{equation}\label{y1}
h(k,\eta_{0})=A\Big(\frac{k}{k_{H}}\Big)^{2+\beta} \coth^{1/2}[\frac{k}{2T}],\;\;\;k\leq k_{E},
\end{equation}

\begin{equation}\label{ke}
h(k ,\eta_{0})=A\Big(\frac{k}{k_{H}}\Big)^{\beta-\gamma} (1+z_{E})^{\frac{-2-\gamma}{\gamma}} \coth^{1/2}[\frac{k}{2T}],\;\;\;k_{E}\leq k\leq k_{H},
\end{equation}

\begin{equation}\label{l}
h(k,\eta_{0})=A\Big(\frac{k}{k_{H}}\Big)^{\beta} (1+z_{E})^{\frac{-2-\gamma}{\gamma}},\;\;\;k_{H}\leq k\leq k_{2},
\end{equation}


\begin{equation}\label{ppp}
h_{m}(k,\eta_{0})=A\Big (\frac{k_{2}}{k_{0}}\Big)^{\beta} \frac{1}{(1+z_{E})^{3}} \Big( \frac{k}{k_{2}} \Big)^{\varrho},\;\;\;k _{2}\leq  k \leq k _{s},
\end{equation}
with
\begin{equation}\label{ss}
\varrho=\frac{\log_{10}(h_{2T})_{k_{s}}-\log_{10}(h_{1T})_{k_{2}}}{\log_{10}(k_{s})-\log_{10}(k_{2})}=\frac{\log_{10}\Big ( \Big(\frac{k_{s}}{k_{2}} \Big)^{1+\beta}  \mathrm {{coth}^{1/2}\Big[\frac{k_{s}}{2T_{*}}\Big]}\Big)}{\log_{10}\Big( \frac{k_{s}}{k_{2}}\Big)},
\end{equation}
where $h_{m}$ is modified amplitude \cite{8} and

\begin{figure}[t]
{\includegraphics[scale=0.45]{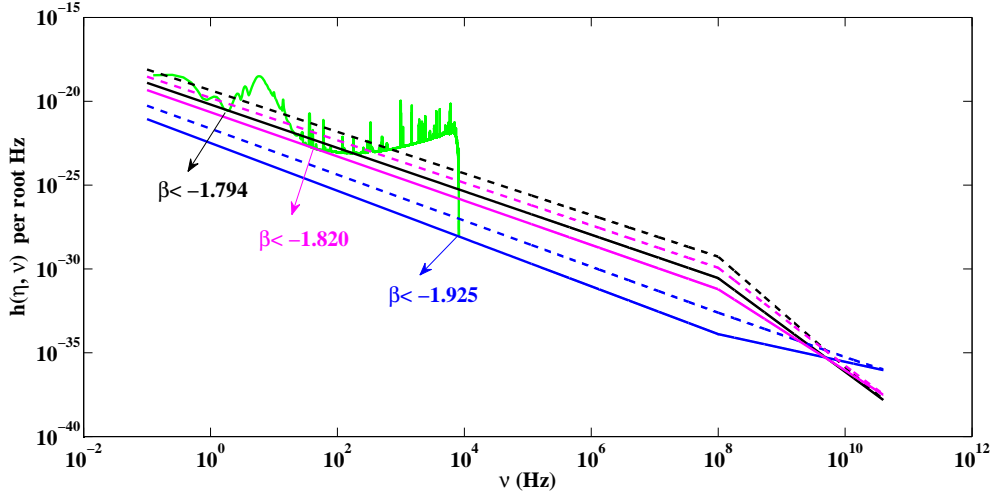}}
\caption{ The spectrum of RGWs in usual case (solid lines) and thermal case (dashed lines)  compared to the Hanford sensitivity (green line).}\label{f1}
\end{figure}
\begin{equation*}
h(k,\eta_{0})=A\Big(\frac{k}{k _{H}}\Big)^{1+\beta-\beta_{s}}\Big(\frac{k _{s}}{k _{H}}\Big)^{\beta_{s}}\Big(\frac{k _{H}}{k _{2}}\Big)(1+z_{E})^{\frac{-2-\gamma}{\gamma}} \coth^{1/2}[\frac{k}{2T_{*}}],
\end{equation*}
\begin{equation}\label{qw}
\;\;\;k _{s}\leq k \leq k _{1}.
\end{equation}
where $T=0.001 \;Mpc^{-1}$\cite{6}, $T_{*}=1.19\times10^{25}\; Mpc^{-1}$ \cite{7}, $\gamma$ is $\Omega_{\Lambda}$ dependent parameter, and $\Omega_{\Lambda}$ is the energy density contrast. We take  the value of redshift $ z_{E}\sim1.3$ and $\gamma\simeq 1.05$ \cite{13} for $\Omega_{\Lambda}=0.692$  from Planck 2015 \cite{1}.
By taking   $k=2\pi\nu$, we can obtain $\nu_{E}=1.93\times10^{-18}$ Hz, $\nu_{H}=2.28\times10^{-18}$ Hz, $\nu_{2}=9.3\times10^{-17}$ Hz, $\nu_{s}=10^{8}$ Hz  and $\nu_1\simeq 4\times10^{10} $ Hz \cite{11,14}.

One can get the constant $A$ without scalar running as follows \cite{15}

\begin{equation}\label{sa}
A=\frac{\Delta_{R}(k_{0})r^{1/2}}{(1+z_{E})^{\frac{-2-\gamma}{\gamma}}}(\frac{\nu_H}{\nu_0})^{\beta},
\end{equation}
where  $\Delta^{2}_{R}(k_{0})$ is the power spectrum of the  curvature perturbation evaluated at the pivot wave number $k^{p}_{0}=k_{0}/a(\eta_{0})=0.002$ Mpc$^{-1}$ \cite{16} with corresponding physical frequency  $\nu_{0}=3.09\times 10^{-18}$ Hz. The $\Delta^{2}_{R}(k_{0})=(2.464\pm0.072\times10^{-9})$ is given by WMAP9$+$eCMB$+$BAO$+$ $H_{0}$ \cite{17}. The tensor to scalar ratio $ r< 0.11 (95 \%CL)$ is based on Planck measurement \cite{18}. We take $r\simeq 0.1$ and also value of  redshift $z_{E}=0.3$  for TT, TE, EE+lowP+lensing  contribution in this work, see appendix.[A] for more details.

\section{The analysis of the spectrum}

 Let us now call the spectrum of the waves in the thermal case  as `thermal spectrum' . In this section we  have supported and upgraded our previous result that obtained in \cite{8} by comparision the  thermal spectrum  with the average measured sensitivity of Adv.LIGO (Hanford and Livingstone) and Dml. We are interested to the  frequency range $\sim($10$^{-1}$Hz -10$^{11}$Hz). Therefore we plotted the spectrum by using eqs.(\ref{ppp}-\ref{qw}) in figs.[\ref{f1}, \ref{as}]. Note that the amplitude rescaled to $h( \nu)/\sqrt{\nu}$ for comparison with Adv.LIGO. The fig.[\ref{f1}] and fig.[\ref{as}] show the spectrum compared to the Hanford (green color) and Livingstone (red color) sensitivity respectively. The solid and dashed lines are stand for the usual and thermal spectrum  respectively. We obtained upper bound on the $\beta$ at three points as a sample in both figures compared to Hanford  and Livingstone  sensitivity. The obtained diagrams tell us, there are  some chances for  detection of the thermal spectrum with modified amplitude in addition to the usual spectrum, see the intersection between the dashed lines and Hanford  and Livingstone  sensitivity in both figures. Also the obtained upper bounds on $\beta$ give us more information about the nature of the evolution of inflation stage.

On the other hand there is a  procedure for  detection based on Dml at GHz band  \cite{5}. The author in \cite{5}  is obtained the sensitivity of the Dml  at the  frequency $4.5$ GHz like:
\begin{figure}[t]
{\includegraphics[scale=0.45]{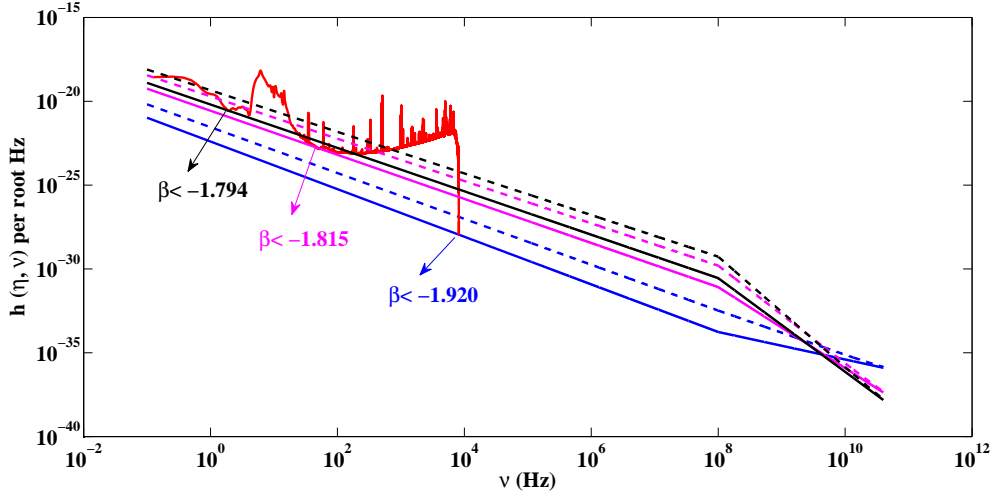}}
\caption{The spectrum of RGWs in usual case (solid lines) and thermal case (dashed lines)  compared to the Livingstone sensitivity (red line).}\label{as}
\end{figure}

\begin{figure}[t]
{\includegraphics[scale=0.45]{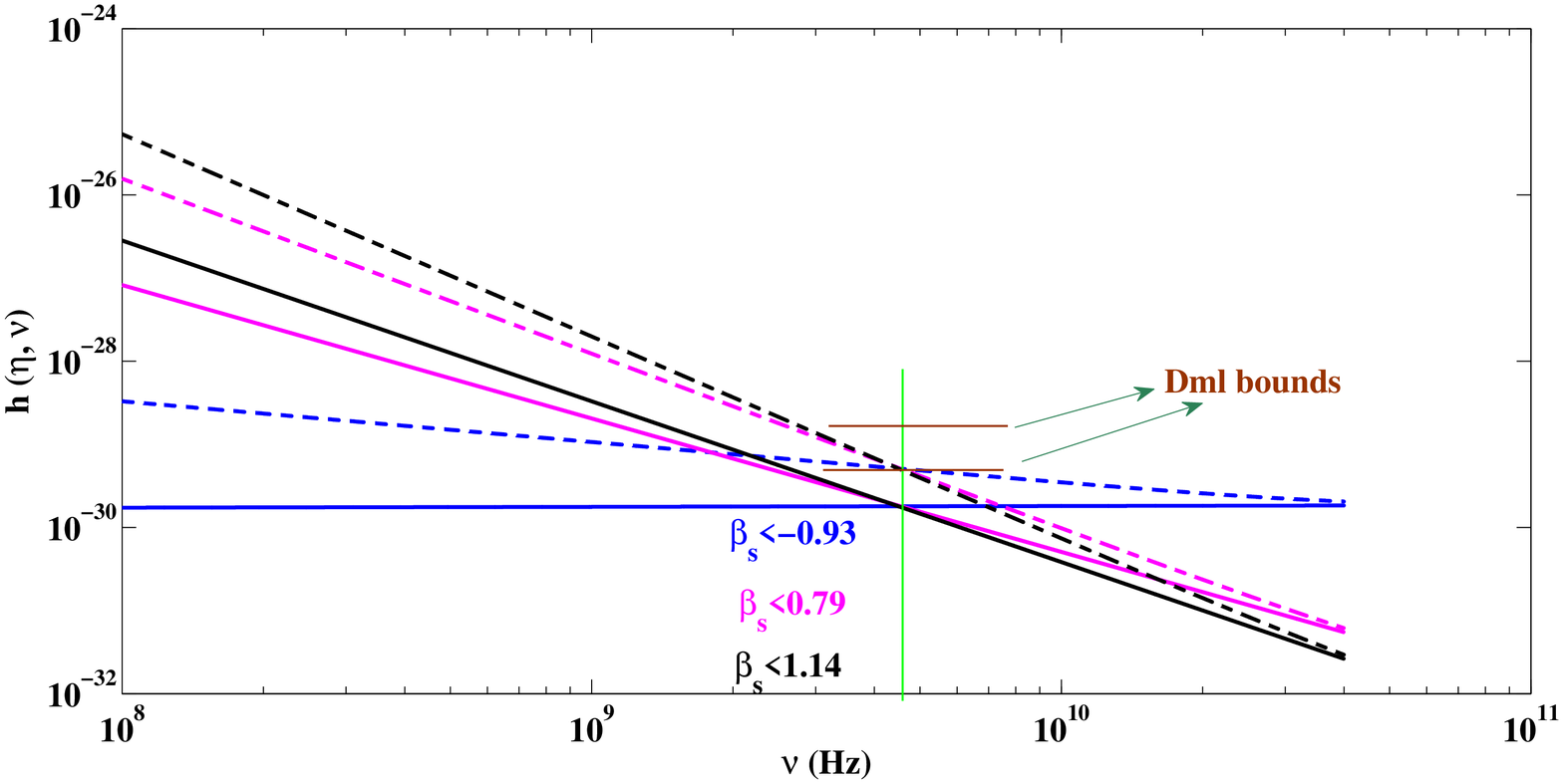}}
\caption{The high frequency range of fig.[\ref{as}] compared to Dml sensitivity at $4.5$ GHz (green vertical line).}\label{qqe}
\end{figure}
\begin{figure}[t]
{\includegraphics[scale=0.45]{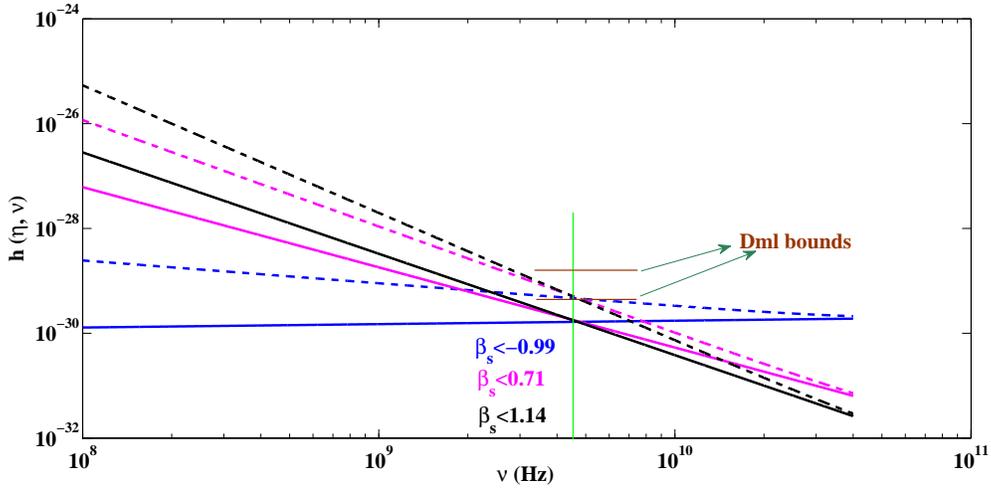}}
\caption{ The high frequency range of fig.[\ref{f1}] compared to Dml sensitivity at $4.5$ GHz (green vertical line).}\label{qe}
\end{figure}

\begin{equation}\label{r}
\sim (5.1\times 10^{-30}-1.6\times 10^{-29}).
\end{equation}

  He is believed that the Dml can not detect the   RGWs due to the gap of $4\sim 5$ orders between the sensitivity of the Dml and the amplitude of the waves. Therefore he is recommended  to upgrade the Dml by using some points that mentioned in \cite{5}. In addition to those points, we claim that the problem of  detection can remove by considering   thermal spectrum. Using eq.(\ref{qw}), we plotted   thermal spectrum (dashed lines) compared to the usual spectrum (solid lines) in the frequency range  $\sim($10$^{8}$Hz -10$^{11}$Hz)   in figs.[\ref{qe}, \ref{qqe}]. The green vertical line stands for the frequency at $4.5$ GHz in both figures. The  upper bounds on $\beta_{s}$ in fig.[\ref{qe}] (fig.[\ref{qqe}]) obtained corresponding to the  upper bounds on $\beta$ in fig.[\ref{f1}] (fig.[\ref{as}]). For given $\beta, \beta_{s}$   the obtained   usual and  thermal amplitude in both figures are $\sim 1.7\times 10^{-30}$ and $\sim 5.1 \times 10^{-30}$  at the frequency $4.5$ GHz  respectively.  Note that for  given $\beta$  and lower $\beta_{s}$ according to the range of $\beta_{s}$ in figs.[\ref{qe},\ref{qqe}], the thermal amplitude will be further. Therefore  based on the range in eq.(\ref{r}) and obtaind usual amplitude  $\sim 1.7\times 10^{-30}$, Dml can not detect the waves in the usual case as the author said in \cite{5}. But by considering the thermal amplitude $\sim 5.1 \times 10^{-30}$, the obtained diagrams tell us that  we are lucky to detect the waves  at the  frequency  $4.5$ GHz. Also the obtained upper bounds on $\beta_{s}$ give us more information about the nature of the evolution of the reheating stage.

\section{Discussion and conclusion }

The RGWs are generated before and during inflation stage.  They contain valuable information about the early universe in the frequency range $\sim($10$^{-19}$Hz -10$^{11}$Hz). The thermal spectrum of RGWs causes the  new amplitude that called `modified amplitude'. The obtained diagrams of the spectrum tell us, there are  some chances for  detection of the thermal spectrum  in addition to the usual spectrum by using Adv.LIGO data of GW150914 and Dml.  Also the $\beta$ and $\beta_s$ have an unique effect on the shape of the spectrum. We found some upper bounds on the $\beta$ and $\beta_s$ by comparison the  usual and thermal spectrum with  Adv.LIGO data of GW150914 and  Dml.
As this result about the upper bounds modified our information about the nature of the evolution of inflation and reheating stages.

\appendix
 \section {}\label{df}

 The general solution of  eq.(\ref{zz}) is a linear combination of the Hankel function with
a generic power law for the scale factor $S=\eta^{u}$ given by

\begin{equation}\label{uy}
f_{k}(\eta)=A_{k}\sqrt{k\eta}H^{(1)}_{u-\frac{1}{2}}(k\eta)+B_{k}\sqrt{k\eta}H^{(2)}_{u-\frac{1}{2}}(k\eta).
\end{equation}

We can write an exact solution $f_{k}(\eta)$ by matching its value and
derivative at the joining points, for of a sequence of successive scale
factors with different $u$ for a given model of the expansion of universe.
The  approximate solution of the spectrum of  RGWs  is usually computed in two limiting cases based on the waves  are   outside ($k^{2} \gg S^{\prime \prime}/S$, short wave   approximation) or inside ($k^{2}  \ll  S^{\prime \prime}/S$, long wave  approximation) of the  barrier. For the RGWs  outside the barrier   the corresponding   amplitude  decrease as $h_k \propto 1/S(\eta) $ while for the  waves inside the barrier,  $h_k = C_k $ simply a constant. Therefore these results can be used to obtain the spectrum for the present stage of  universe \cite{11,s}.
The history of   expansion of the universe can written as follows:

\textbf{a}) Inflation stage:
\begin{equation}\label{p}
S(\eta)=l_{0}|\eta |^{1+\beta},\;\;\;\;\;\;-\infty <\eta\leq \eta_{1},
\end{equation}
where $1+\beta <0$, $\eta<0$ and $l_{0}$ is a constant.

\textbf{b}) Reheating stage:

\begin{equation}\label{pq}
S(\eta)=S_{z}(\eta - \eta_{p})^{1+\beta_{s}},\;\;\;\;\;\;\eta_{1} <\eta\leq \eta_{s},
\end{equation}
where $1+\beta_{s}>0$, see for more details \cite{11}.

\textbf{c}) Radiation-dominated stage:
\begin{equation}
S(\eta)=S_{e}(\eta-\eta_{e}),\;\;\;\;\;\;\eta_{s}\leq \eta \leq \eta_{2}.
\end{equation}

\textbf{d}) Matter-dominated stage:
\begin{equation}
S(\eta)=S_{m}(\eta-\eta_{m})^{2},\;\;\;\;\;\;\eta_{2}\leq \eta \leq \eta_{E},
\end{equation}
where $\eta_{E}$ is the time when the dark energy density $\rho_{\Lambda}$ is equal to the matter energy density $\rho_{m}$.
The value of  redshift $z_{E}$ at  $\eta_{E}$ is  $(1+z_{E})=S(\eta_{0})/S(\eta_{E})\sim1.3$ for TT, TE, EE+lowP+lensing  contribution based on Planck 2015 \cite{1}
 where $\eta_{0}$ is the present time.

\textbf{e}) Accelerating stage:
\begin{equation}\label{1w}
S(\eta)=\ell_{0}|\eta- \eta_{a} |^{-\gamma},\;\;\;\;\;\;\eta_{E}\leq \eta \leq\eta_{0},
\end{equation}

For normalization purpose of $S$, we put $|\eta_{0}-\eta_{a}|=1$  which fixes the  $\eta_{a}$, and the constant $\ell_{0}$ is fixed by the following relation,
\begin{equation}
\frac{\gamma}{H_{0}}\equiv \Big(\frac{S^{2}}{S^{\prime}}\Big)_{\eta_{0}}=\ell_{0},
\end{equation}
where $\ell_{0}$ is  the Hubble radius at present with $H_{0}=67.8$ km s$^{-1}$Mpc$^{-1}$ from Planck 2015 \cite{1}.
 The  wave number $k_{H}$ corresponding to the present Hubble radius is
$k_{H}=2\pi S(\eta_{0})/ \ell_{0}=2\pi \gamma$.
There is another wave number
$k_{E}=\frac{2\pi S(\eta_{E})}{1/H_{0}}=\frac{k_{H}}{1+z_{E}},$
that its  wavelength at the time $\eta_{E}$ is the Hubble radius $1/H_{0}$.
By matching $S$ and $S^{\prime}/S$ at the joint points, one gets
\begin{equation}\label{kk}
l_{0}=\ell_{0}b\zeta_{E}^{-(2+\beta)}\zeta_{2}^{\frac{\beta-1}{2}}\zeta_{s}^{\beta}\zeta_{1}^{\frac{\beta-\beta_{s}}{1+\beta_{s}}},
\end{equation}
where $b\equiv|1+\beta|^{-(2+\beta)}$, $\zeta_{E}\equiv\frac{S(\eta_{0})}{S(\eta_{E})}$, $\zeta_{2}\equiv\frac{S(\eta_{E})}{S(\eta_{2})}$, $\zeta_{s}\equiv\frac{S(\eta_{2})}{S(\eta_{s})}$, and $\zeta_{1}\equiv\frac{S(\eta_{s})}{S(\eta_{1})}$.
 With these specifications, the functions $S(\eta)$ and $S^{\prime}(\eta)/S(\eta)$ are fully determined \cite{11,15}.

The power spectrum  of  RGWs is defined  as
\begin{equation}\label{pow}
\int_0 ^\infty h^2 (k,\eta) \frac{dk} {k} = \langle 0 | h^{ij}({\bf x},\eta) h_{ij}({\bf x},\eta) |0 \rangle .
\end{equation}
 Substituting eq.(\ref{1}) in eq.(\ref{pow}) with same contribution of each polarization, we get
 \begin{equation}\label{pp}
 h(k,\eta)= \frac{4 l_{pl}}{\sqrt{\pi}} k | h(\eta)|.
 \end{equation}
The spectrum at the present time $ h(k,\eta_0)$ can be obtained, provided the initial  spectrum is specified. 
The  initial amplitude of the spectrum is given by
\begin{equation}\label{bet}
h(k,\eta_i)= A{\left(\frac {k}{k_H}\right)}^{2+\beta},
\end{equation}
where the constant $A$  can be determined by quantum normalization \cite{11,15}.

  Therefore the amplitude  of the spectrum   for  different ranges of wave numbers are given by \cite{11,12,15}

\begin{equation}\label{y}
h(k,\eta_{0})=A\Big(\frac{k}{k_{H}}\Big)^{2+\beta},\;\;\;k\leq k_{E},
\end{equation}

\begin{equation}\label{ke}
h(k ,\eta_{0})=A\Big(\frac{k}{k_{H}}\Big)^{\beta-\gamma} (1+z_{E})^{\frac{-2-\gamma}{\gamma}},\;\;\;k_{E}\leq k\leq k_{H},
\end{equation}

\begin{equation}\label{l}
h(k,\eta_{0})=A\Big(\frac{k}{k_{H}}\Big)^{\beta} (1+z_{E})^{\frac{-2-\gamma}{\gamma}},\;\;\;k_{H}\leq k\leq k_{2},
\end{equation}

\begin{equation}\label{o}
h(k,\eta_{0})=A\Big(\frac{k}{k _{H}}\Big)^{1+\beta}\Big(\frac{k_{H}}{k_{2}}\Big)(1+z_{E})^{\frac{-2-\gamma}{\gamma}},\;\;\;k _{2}\leq  k \leq k _{s},
\end{equation}

\begin{equation*}\label{oo}
h(k,\eta_{0})=A\Big(\frac{k}{k _{H}}\Big)^{1+\beta-\beta_{s}}\Big(\frac{k _{s}}{k _{H}}\Big)^{\beta_{s}}\Big(\frac{k _{H}}{k _{2}}\Big)(1+z_{E})^{\frac{-2-\gamma}{\gamma}},
\end{equation*}
\begin{equation}
\;\;\;k _{s}\leq k \leq k _{1}.
\end{equation}

 \section {}

 An effective approach to deals with the thermal  vacuum state is the thermo-field dynamics (TFD)\cite{34}. In this approach  a tilde space  is needed besides the usual
Hilbert space, and the direct product space is made up of the these two spaces. Every operator and state in the Hilbert space has the corresponding counter part in the tilde
space \cite{34}. Therefore a  thermal  vacuum state  can be  defined as
\begin{equation}\label{16}
|Tv\rangle ={\cal T }(\theta_{k})|0\; \tilde{0}\rangle,
\end{equation}
where
 \begin{equation}\label{333}
{\cal T }(\theta_{k})=\mathrm{exp} [-\theta_{k} (a_{\bf {k}}\tilde{a}_{\bf {k}}-a_{\bf{k}}^{\dagger}\tilde{a}_{\bf {k}}^{\dagger})],
\end{equation}
is the thermal operator and $|0\; \tilde{0}\rangle$ is the two mode vacuum state at zero temperature. The  quantity $\theta_{k}$ is  related to   the average number of the thermal particle, $\bar{n}_{k}=\mathrm{sinh}^{2}\theta_{k}$. The $\bar{n}_{k}$  for given  temperature T is
provided   by the Bose-Einstein distribution
$\bar{n}_{k}=[\mathrm{exp}( k /T)-1]^{-1}$,
where $\omega_{k}$ is the resonance frequency of the field, see Ref.\cite{8} for more details.

\end{document}